\documentclass[%
 reprint,
superscriptaddress,
 amsmath,amssymb,
 aps,
]{revtex4-2}
\usepackage{graphicx}
\usepackage{dcolumn}
\usepackage{color}
\usepackage{comment}
\usepackage{multirow}
\usepackage{hyperref}
\usepackage{float}
\makeatletter
\let\newfloat\newfloat@ltx
\makeatother
\usepackage{algorithm}
\usepackage{algpseudocode}
\usepackage{bm}
\usepackage{mathrsfs}
\usepackage{float}
\usepackage{amsmath}
\usepackage{warpcol}
\graphicspath{{Images/}}
\usepackage{xcolor}
\definecolor{red1}{rgb}{0.6,0,0}

\begin{document}

\title{Design of 2D Skyrmionic Metamaterials Through Controlled Assembly}

\newcommand{\KTH}{Department of Applied Physics, School of Engineering Sciences, KTH Royal Institute of Technology, 
AlbaNova University Center, SE-10691 Stockholm, Sweden}

\newcommand{\KTHeecs}{Division of Computational Science and Technology, School of Electrical Engineering and Computer Science, KTH Royal Institute of Technology, AlbaNova University Center, SE-10691 Stockholm, Sweden}

\newcommand{\WISEKTH}{Wallenberg Initiative Materials Science for Sustainability (WISE), KTH Royal Institute of Technology, SE-10044 Stockholm, Sweden}
\newcommand{\SeRC}{Swedish e-Science Research Center, KTH Royal Institute of Technology, SE-10044 Stockholm, Sweden}
\newcommand{\SC}{College of Energy, Soochow Institute for Energy and Materials InnovationS (SIEMIS), and Jiangsu Provincial Key Laboratory for Advanced Carbon Materials and Wearable Energy Technologies, Soochow University, Suzhou, 215006, China}
\newcommand{\Uppsala}{Department of Physics and Astronomy, Uppsala University, Box 516, SE-75120 Uppsala, Sweden}
\newcommand{\WISEUU}{Wallenberg Initiative Materials Science for Sustainability (WISE), Uppsala University, Box 516, SE-75120 Uppsala, Sweden}
\newcommand{\Orebro}{School of Science and Technology, \"Orebro University, SE-701 82, \"Orebro, Sweden}
\newcommand{\Stockholm}{Department of Materials and Environmental Chemistry, Stockholm University, SE-10691 Stockholm, Sweden}
\newcommand{\UppsalaChem}{Department of Chemistry - Ångström Laboratory, Uppsala University, Box 538, Uppsala, SE-751 21, Sweden}
\newcommand{\CASszbio}{Key Laboratory of Quantitative Synthetic Biology, Shenzhen Institute of Synthetic Biology, Shenzhen Institutes of Advanced Technology, Chinese Academy of Sciences, Shenzhen 518055, China}
\newcommand{\Kalmar}{Department of Physics and Electrical Engineering, Linnaeus University, SE-39231 Kalmar, Sweden}

\author{Qichen Xu*}
    \affiliation{\Uppsala}
    \affiliation{\KTH}
    \affiliation{\SeRC}
    \thanks{qichenx@kth.se}
\author{Zhuanglin Shen†}
    \affiliation{\CASszbio}
    
\author{Alexander Edström†}  
    \affiliation{\KTH}
    \affiliation{\SeRC}
\author{I. P. Miranda†}
\affiliation{\Uppsala}
\affiliation{\Kalmar}
\author{Zhiwei Lu}
    \affiliation{\KTH}
\author{Anders Bergman}
     \affiliation{\Uppsala}
\author{Danny Thonig}
     \affiliation{\Orebro}
     \affiliation{\Uppsala}
\author{Wanjian Yin}
    \affiliation{\SC}
\author{Olle Eriksson}
     \affiliation{\Uppsala} 
         \affiliation{\WISEUU}
\author{Anna Delin*}
    \affiliation{\KTH}
    \affiliation{\SeRC}
    \affiliation{\WISEKTH}
    \thanks{annadel@kth.se}
    \thanks{† These authors contributed equally}

\date{\today}

\begin{abstract}
Despite extensive research on magnetic skyrmions and antiskyrmions, a significant challenge remains in crafting nontrivial high-order skyrmionic textures with varying, or even tailor-made, topologies. We address this challenge, by focusing on a construction pathway of skyrmionic metamaterials within a monolayer thin film and suggest several skyrmionic metamaterials that are surprisingly stable, i.e., long-lived, due to a self-stabilization mechanism. This makes these new textures promising for applications. Central to our approach is the concept of 'simulated controlled assembly', in short, a protocol inspired by 'click chemistry' that allows for positioning topological magnetic structures where one likes, and then allowing for energy minimization to elucidate the stability. Utilizing high-throughput atomistic-spin-dynamic simulations alongside state-of-the-art AI-driven tools, we have isolated skyrmions (topological charge $Q=1$), antiskyrmions ($Q=-1$), and skyrmionium ($Q=0$). These entities serve as foundational 'skyrmionic building blocks' to form the here reported intricate textures.  In this work, two key contributions are introduced to the field of skyrmionic systems. First, we present a a novel combination of atomistic spin dynamics simulations and controlled assembly protocols for the stabilization and investigation of new topological magnets. Second, using the aforementioned methods we report on the discovery of skyrmionic metamaterials.
\end{abstract}

\maketitle

Skyrmionic textures represent swirling topological spin configurations, typically stabilized by the Dzyaloshinskii–Moriya interaction (DMI). There are three primary skyrmionic textures: skyrmions \cite{generate_sk_1,generate_sk_2,generate_sk_3,casiraghi2019individual,zeissler2020diameter,wang2022mechanical}, antiskyrmions \cite{ask_create_move_1,ask_create_move_2,ask_create_move_3,ask_create_move_4,Ritzmann2018,ask_create_move_6}, and skyrmionium\cite{SKM_create_move_1,SKM_create_move_2,SKM_create_move_3}, distinguished by their topological charges, morphology and properties (such as the presence or not of the skyrmion Hall effect \cite{Chen2017}).
Moreover, recent studies have identified complex, higher-order skyrmionic textures carrying an absolute topological charge greater than 1\cite{xu2023metaheuristic,foster2019two,desplat2019paths}. The emergent physical properties of skyrmionic textures show great potential in the design of next-generation sustainable spintronic applications, such as neuromorphic computing\cite{song2020skyrmion}, race-track memory\cite{fert2017magnetic}, data encoding, and storage\cite{tokura2020magnetic}.

Advances in understanding interactions among skyrmionic textures, such as attraction, repulsion, and annihilation, resonate with the well-established principles of 'Click chemistry' \cite{kolb2001click}.  'Click chemistry' is a process characterized by the synthesis of substances through the union of smaller units like molecules, with a focus on controlled assembly \cite{lee2021controlled,ishiwari2018supramolecular,ten2016two,li2020smart,tierno2014recent,whitesides2002self,grzelczak2010directed,macfarlane2021emerging,levin2020biomimetic,penfold2019emerging,das2021chemically}. Expanding this concept to 'click magnetism', we introduce 'skyrmionic controlled assembly', utilizing skyrmions, antiskyrmions, and skyrmioniums as elemental skyrmionic building blocks. The simulated controlled assembly protocol involves manually manipulating the position of these skyrmionic building blocks, or known textures, to induce various assembly processes that give origin to new textures with arbitrary topological charges. This methodology not only facilitates the integration of textures, but also enables the creation of metastable extended magnetic textures from these building blocks.

These extended textures are similar to a class of artificial materials, namely metamaterials\cite{hess2012active,zhao2023topological}, which are defined by their custom-engineered configurations and bespoke building blocks, e.g., nanoscale wrinkles and split-ring resonators. Metamaterials are extensively applied in nanostructure designs in the fields of optics and plasmonics to develop sustainable applications which offer capabilities beyond those of conventional materials, such as a negative refractive index\cite{zheludev2012metamaterials}. 
In this work, similar to the definition of metamaterials, we propose a skyrmionic metamaterial concept designed to accelerate the discovery of novel promising magnetic textures with intricate topologies for spintronics applications. We elaborate on this concept by detailing the creation of \textit{i)} lattice-like skyrmionic metamaterials, defined as spin-texture lattices that possess a topological charge greater than 1 within their primitive unit cells, as well as \textit{ii)} flake-like skyrmionic metamaterials, which are large-scale high-order antiskyrmions with an underlying lattice structure, and \textit{iii)} cell-like skyrmionic metamaterials that incorporate a nesting of various texture types. Each of these magnetic textures showcases a non-trivial topology and is constructed from fundamental skyrmionic units. Based on the aforementioned 'Click chemistry' notion applied to skyrmionic textures, the key concept of our systematic exploration of skyrmionic metamaterials is also the controlled assembly, which enables the artificial placement of these elemental skyrmionic building blocks.

Employing spin dynamics simulation tools in our controlled assembly protocols, we investigate the relative stability of the skyrmionic metamaterials. Specifically, we perform high-throughput atomistic-spin-dynamic (ASD) simulations, using the UppASD package\cite{UppASD_book}. The material class we propose consists of a monolayer 
film, which inherently supports non-trivial topologies\cite{xu2023metaheuristic,zhang2017skyrmion,tang2021magnetic,kind2020existence,rybakov2019chiral,kuchkin2020magnetic}. We present a theoretical
pathway for constructing various complex topological textures from 
skyrmionic building-blocks, including reported high-order antiskyrmions and skyrmion bags, as well as the aforementioned skyrmionic metamaterials.

\begin{figure*}
    \centering
    \includegraphics[width=16cm]{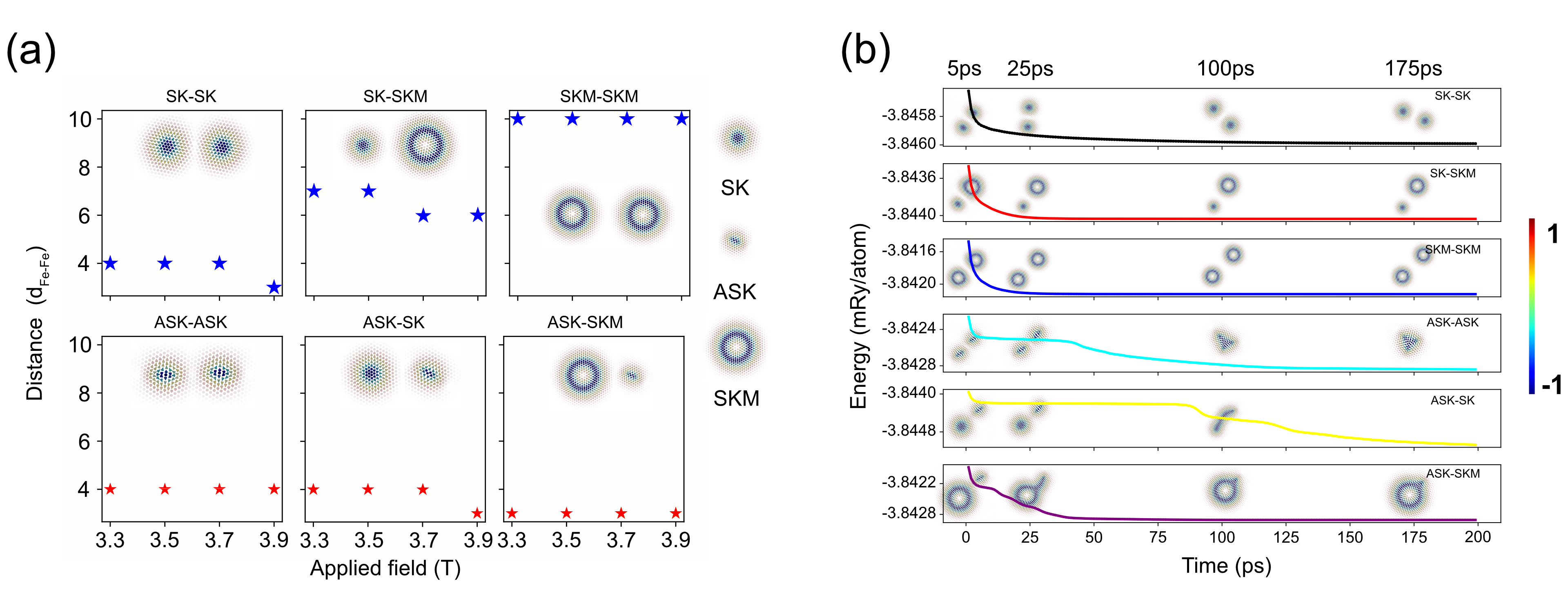}
    \caption{\textbf{Pair-wise interaction between topological building blocks.} (a) Pictorial representations of pair interaction dynamics as a function of applied magnetic field, for all pair-combinations of skyrmions, antiskyrmions, and skyrmionium. For comprehensive visualizations, we refer to the supplementary movies S1-S4. The red and blue stars indicate the calculated equilibrium (stationary) distances for each pair as influenced by the external magnetic field, with distances expressed in units of the Fe-Fe equilibrium distance $d_{\rm Fe-Fe} = 0.272$\,nm.
    Blue stars indicate that for shorter distances the pair interaction energy is repulsive (equilibrium distance). 
    Red stars, conversely, denote that for shorter distance, the spin textures either merge or annihilate (critical distance).
    (b) Temporal evolution of the total energy for each pair assembly process, as derived from ASD simulations. The $y$-axis quantifies the total energy. Accompanying subplots provide snapshots of the spin configurations during the assembly process.
    Color scheme: The adjoining colorbar clarifies the spin orientation in these configurations: dark blue color (-1) signifies spins pointing downwards, away from the viewer, while red color (1) represents spins pointing upwards, towards the viewer. Whit color indicates that the spin is parallel to the plane.
    }
        \label{fig:fig1}
    \end{figure*}

\begin{figure*}
    \centering
    \includegraphics[width=16cm]{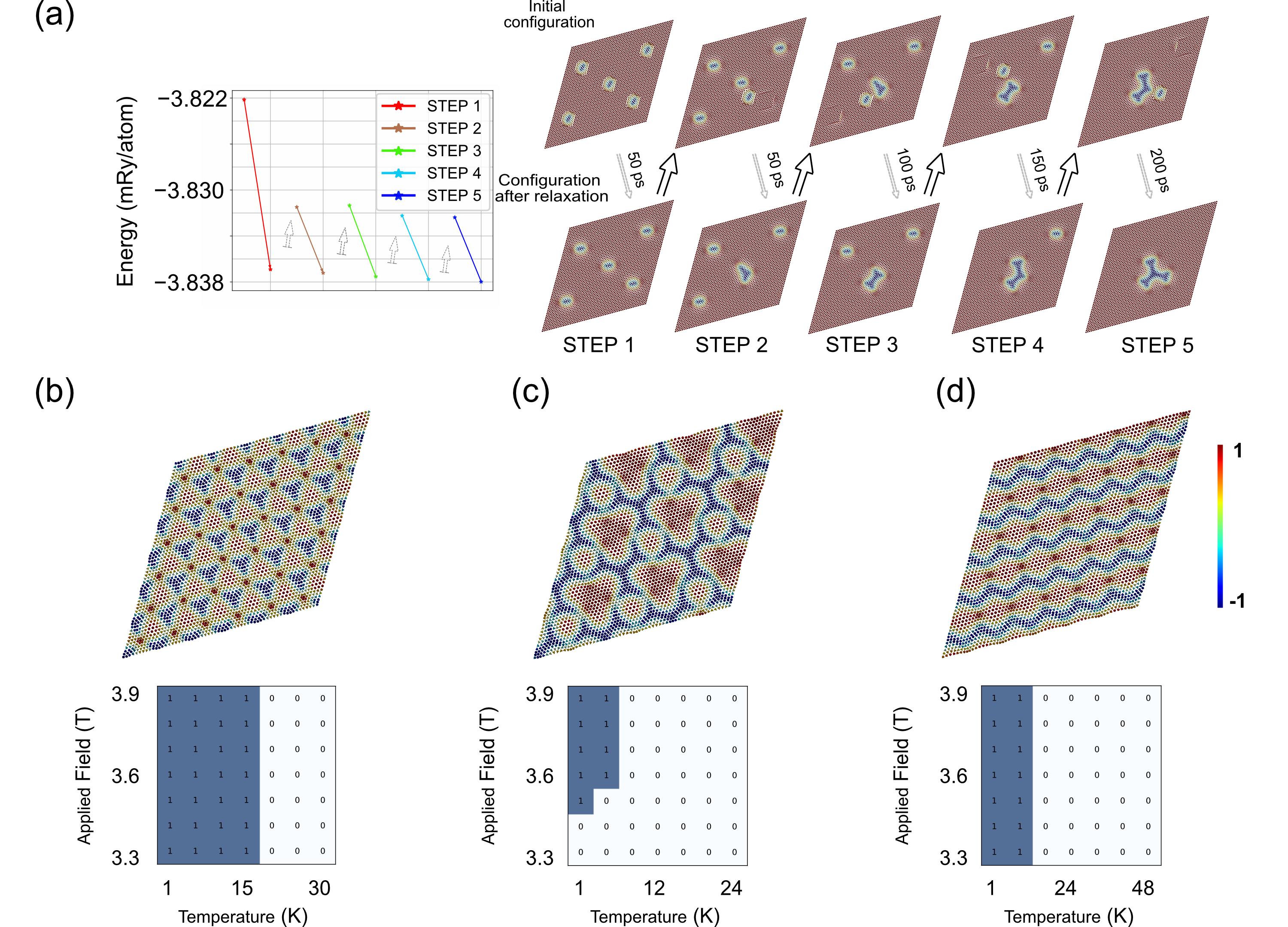}
    \caption{\textbf{Construction and stability of lattice-like skyrmionic metamaterials.}
    Construction of an individual high-order antiskyrmion (the prefabricated building block) and subsequent construction of extended skyrmionic periodic textures (also called "lattice-like skyrmionic metamaterials" in the following) by combining many individual high-order antiskyrmions. The ASD simulation system size is $60 \times 60 $ spins in all subfigures.
    The color scheme is the same as in Fig.\,1.
    (a) Illustration of the step-by-step formation of a triangular high-order antiskyrmion via controlled assembly. The left diagram presents the initial- and post-relaxation energies at each stage, with relaxation times increasing from 50 ps (step 1) to 200 ps (step 5). Top and bottom subfigures display the corresponding initial and postrelaxation spin configurations. Additional details regarding the step-by-step formation of a high-order triangular antiskyrmion are available in Supplementary Movie S5.
    (b-d) Three distinct examples of lattice-like skyrmionic metamaterials. Here,  the rotation of each prefabricated building block is hindered due to their position within an extended texture and implied interaction with adjacent prefabricated building blocks. 
    In the top panels of figures (b-d), the lattice-like skyrmionic metamaterials in real space are shown. The bottom panels each display a stability map analysis of the corresponding lattice-like skyrmionic metamaterial as a function of applied magnetic field and temperature. A value of 1 indicates stability after 200 ps ASD simulation, while 0 denotes lattice disintegration occuring at some point before the 200 ps mark.
    (b) 
    lattice-like skyrmionic metamaterial formed from $Q=-2$ high-order antiskyrmions. (c) A lattice-like skyrmionic metamaterial formed from $Q=-5$ high-order antiskyrmions.
    (d) An lattice-like skyrmionic metamaterial composed of $Q=-12$ high-order antiskyrmions. 
    }
    \label{fig:fig2}
\end{figure*}

\begin{figure*}
    \centering
    \includegraphics[width=16cm]{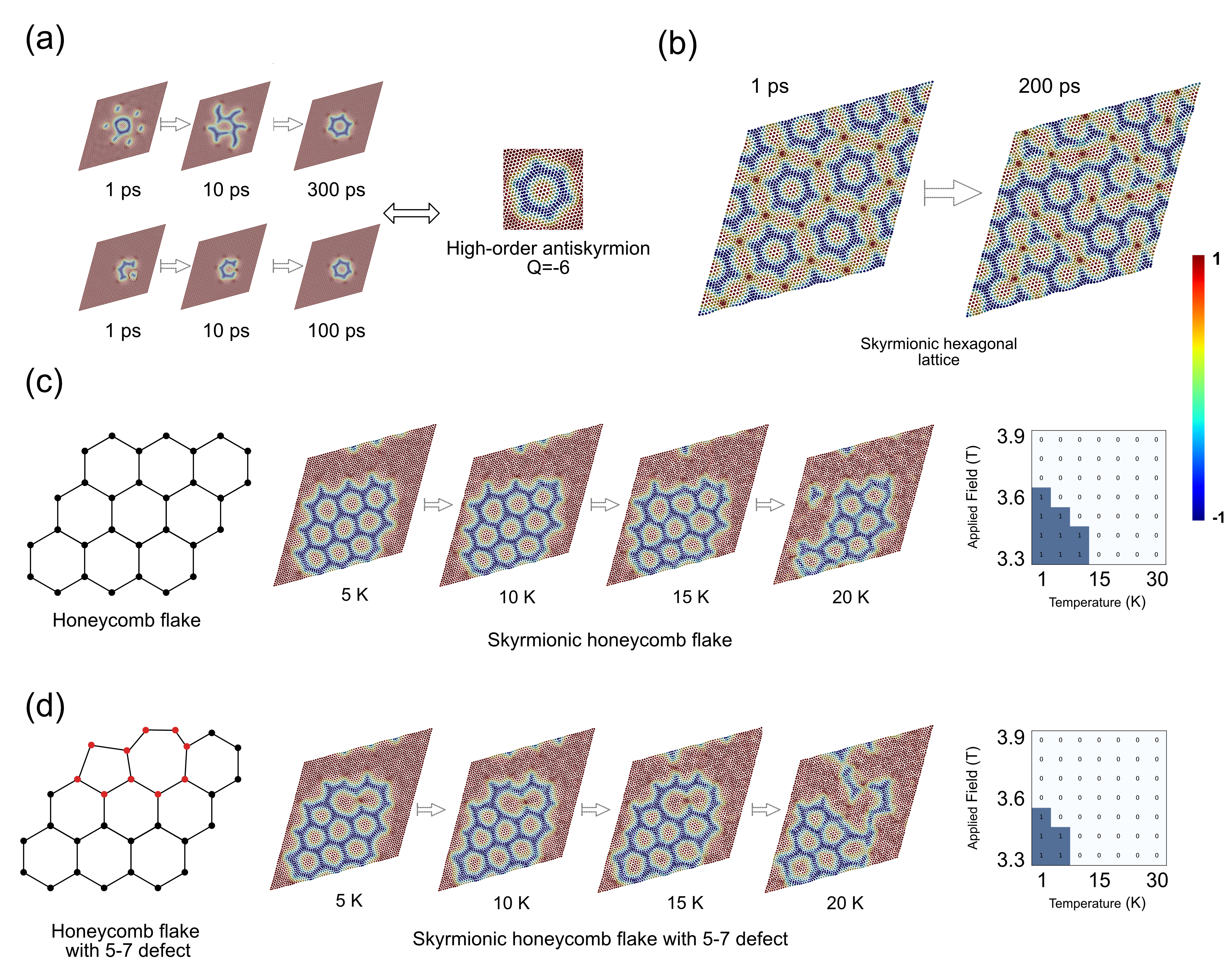}
    \caption{\textbf{Building process and stability of flake-like skyrmionic metamaterials.} 
     (a) Two distinct methodologies for assembling a high-order ring-shaped antiskyrmion with 
     $Q=-6$.
     The top pathway involves combining six antiskyrmions with one skyrmionium, whereas the bottom pathway utilizes one high-order antiskyrmion 
     ($Q=-5$)
     paired with an additional antiskyrmion. Additional details regarding the formation of a high-order ring-shaped antiskyrmion are available in Supplementary Movie S6.
    (b) Temporal progression of a skyrmionic honeycomb lattice at 1\,K. The left image depicts the initial spin configuration, while the right image illustrates the state after 200\,ps ASD time evolution.
    (c) and (d) The left-most panel presents a flawless honeycomb flake-like skyrmionic metamaterial alongside a variant with a 5-7 ring defect (indicated by red dots). The central subplots in (c) and (d) display the real-space spin configurations of a skyrmionic honeycomb flake after 200\,ps ASD simulations at four distinct temperatures. The far-right panels in (c) and (d) showcase the respective stability phase diagrams. Here, a value of 1 indicates stability after 200\,ps of ASD simulations, while 0 denotes that the flake vanished.
    The color scheme in the spin configurations is the same as in Fig. 1.
}
    \label{fig:fig3}
\end{figure*}

\section{Results}
\subsection{Local assembly and interaction profiles}

\label{sec:local-assembly}
For the design of controlled assembly simulations in a 2D skyrmionic system, it is imperative to first analyze pairwise assembly, the relative stability of individual topological objects, and the interactions between the skyrmionic building blocks (i.e., between pairs of skyrmions, antiskyrmions, and skyrmioniums in all combinations). An efficient approach to study these behaviors involves examining the stationary configurations that emerge from performing spin dynamics simulations. There are two cases: some pairs attain a local energy minimum retaining a finite distance between them. For other pairs, the constituents will repel each other for large distances, but will merge (or annihilate) when forced closer to each other than a certain critical distance. 

In our previous studies \cite{xu2023metaheuristic,xu2022genetic}, we observed that Pd/Fe/Ir(111) can host a variety of high-order, small-sized skyrmionic textures, especially close to the transition region between skyrmion-lattice and ferromagnetic (single-domain) ground states. This agrees with several theoretical investigations that independently reported both antiskyrmions and higher-order topological spin configurations as metastable states in this system \cite{vonMalottki2017,Nagyfalusi2024,Ritzmann2020,Goerzen2023,Dupe2016,Ritzmann2018}. In contrast to other systems that can host both skyrmions and antiskyrmions, such as Mn$_{1.4}$Pt$_{0.9}$Pd$_{0.1}$Sn \cite{Jena2020}, these small textures in Pd/Fe/Ir(111) are highly advantageous for the development of next-generation, high-information-density spintronics devices. Moreover, such sizes allow for a fully atomistic modeling with long-range magnetic interactions with \textit{ab-initio} accuracy \cite{miranda2022band}. For these reasons, in the present paper, we focus our investigations on this system -- but, as we discuss in subsequent sections, the ideas/results presented here may be extended to other frustrated materials; a detailed discussion of the generalization of results can be found in Section \ref{sec:Generalization}.

Unless otherwise mentioned, the Gilbert damping parameter in the ASD is $\alpha=0.1$ and we do all the simulations in a $B^{\textnormal{ext}}=3.5$ T magnetic field. We suggest this value of damping as a good lower-bound estimate for the $\alpha$ parameter of Pd/Fe/Ir(111), given the typical values for $3d$ ferromagnets/heavy-metal heterostructures \cite{Barati2014}, and the enhancement influenced by the reduced thickness (single Fe monolayer) \cite{Tserkovnyak2002} and the presence of noncollinearity \cite{Rozsa2018}. For the chosen value of $B^{\textnormal{ext}}$, the ground-state is ferromagnetic (single-domain). 

To minimize thermal effects, not central to the study in this section, we set the simulation temperature to 1 mK. Generally, with fixed interaction parameters, a stronger applied field reduces both equilibrium and critical distances for combinations of skyrmions, antiskyrmions, and skyrmioniums, as shown in Figure \ref{fig:fig1} (a).

As shown in the top three subfigures of Figure \ref{fig:fig1}(a), we observe distinct behaviors. The combination of skyrmion and skyrmionium (middle figure, top panel of Fig.1a) exhibits isotropic outer boundaries, resulting in rotation-invariant repulsion. However, the dynamics between the different systems presented in Fig.1 differ. For instance, in skyrmion-skyrmion pairs, they separate until achieving an equilibrium distance through gyroscopic rotational dynamics\cite{capic2020skyrmionmacfarlane2021emerging}. 

In contrast, with skyrmion-skyrmionium pairs, only the skyrmionium moves while the skyrmion remains stationary. When an antiskyrmion, characterized by an anisotropic boundary, is part of the skyrmionic pair, the interaction behavior changes. Upon approaching the critical distance from far away
, these components exhibit attraction, leading to assembly\cite{PhysRevB.99.174409}. As depicted in the bottom three subfigures of Figure \ref{fig:fig1}(a), two skyrmionic textures assemble when their initial gap
is smaller than the critical distance, resulting in either a high-order antiskyrmion or a vanishing ferromagnetic state. During the assembly of two antiskyrmions, as illustrated in Figure \ref{fig:fig1}(b), a metastable "blocking point" emerges. 

This object is in essence a quasi-equilibrium state that appears when two antiskyrmionic textures are close enough and the self-rotation is prevented in some way.
Once a large enough lattice-like skyrmionic metamaterial is constructed, self-rotation becomes very difficult, since this would require the entire metamaterial to rotate. The result is that the lattice-like skyrmionic metamaterial becomes a very long-lived metastable state, despite its high complexity.

For a further discussion of the blocking-point concept, see Supplementary Note 2.

\subsection{High-order antiskyrmion lattice}

Three examples of self-stabilized lattice-like skyrmionic metamaterials that we have discovered are shown in the top subfigures of Figure \ref{fig:fig2}(b),(c), and (d). By using numerically  prefabricated building blocks (see text below), i.e., high-order antiskyrmions with $Q=-2$, $-5$, and $-12$, it is possible to construct such skyrmionic metamaterials, essentially similar to, e.g., artificial spin-ice structures and lattice based metamaterials. Here, high-order antiskyrmions represent magnetic textures with topological charge $Q<-1$ \cite{xu2023metaheuristic,sen2022manipulation}. To quantify the stability of the skyrmionic metamaterials of Fig.\ref{fig:fig2}, we computed stability maps with temperature and external field as critical parameters. These stability maps are depicted at the bottom of Fig.\ref{fig:fig2}. Note that in these figures stability is marked as a blue region (also marked with a 1) while the white region, marked with a 0, denotes an unstable configuration, 
that anihilates within the simulation time. In this stability analysis we terminated the simulations after 200 ps simulation time, and simply inspected if a configuration was stable by monitoring that it reached a static state. From this analysis, we can draw the conclusion that the magnetic configurations are long-lived (compared to, e.g., typical spin precession and relaxation timescales), with a lifetime exceeding 200 ps. Interestingly, the stability analysis of Fig.\ref{fig:fig2} shows that lattices of higher order skyrmions can survive significant temperature and field ranges. In this regard, it should be noted that a detailed analysis of the potential energy surface using geodesic nudged elastic band (GNEB) \cite{Bessarab2015} could better quantify the lifetimes and transition energy barriers of the formed structures. However, applying GNEB to high-order topological structures in frustrated magnets remains challenging and will be explored in future work.

Below, we detail our approach for constructing and discovering long-lived lattice-like skyrmionic metamaterials. In essence, the first step is to construct high-order antiskyrmions (referred to as prefabricated building blocks henceforth) from simple antiskyrmions. Then, these prefabricated building blocks are combined together in such a way that self-rotation is prevented, which in turn stabilizes the resulting lattice-like skyrmionic metamaterial.

In this way we start by using single antiskyrmions to construct prefabricated building blocks via simulated controlled assembly. This prefabrication process is guided by the site-based topological charge distribution rule discussed in Section  \ref{sec:method} and the fact that antiskyrmions show an attractive interaction with other structures (skyrmion, skyrmionium, antiskyrmion), as demonstrated in Section \ref{sec:local-assembly}. 
A subtle issue in the prefabrication process is how to place the antiskyrmions so that they assemble to the building block we want to fabricate. In the simulation, this is done by placing them at the appropriate distances from each other. 

In an experimental setting, this may be realized using, for example, gradient magnetic fields \cite{zhang2018manipulation,casiraghi2019individual,peng2022formation,huang2017stabilization,koshibae2014creation}, electric fields created by the scanning tunneling microscope \cite{Wieser2017}, or even temperature gradients \cite{Kong2013}. 
In Figure \ref{fig:fig2}(a) we give one example on how to fabricate a $Q=-5$ high-order antiskyrmion, with five controlled-assembly steps, starting from five isolated antiskyrmions (one at the center and four at the edges). During the simulation, we manually moved isolated antiskyrmions from the edges to the center and relaxed the spin structure to ensure the successful formation of the new high-order antiskyrmion and the completion of the fabrication process. Finally, the triangle-shaped high-order antiskyrmion ($Q=-5$) is fabricated by controlled placement of a single antiskyrmion. 
The energy change with each step is illustrated in the left part of Figure \ref{fig:fig2}(a). This result indicates a dynamic construction method for building high-order antiskyrmions: sequentially manipulating pregenerated antiskyrmions.
Once we have constructed the prefabricated building blocks, these can be used to construct lattice-like skyrmionic metamaterials. 
Numerous structures are of course possible, and which lattice-like skyrmionic metamaterial will result depends on the relative placement of the prefabricated building blocks. For instance, in order to create a lattice-like skyrmionic metamaterial with a hexagonal lattice as shown in Figure \ref{fig:fig2}(b), the prefabricated building blocks are placed so that they form a hexagonal lattice.

\begin{figure*}
    \centering
    \includegraphics[width=15cm]{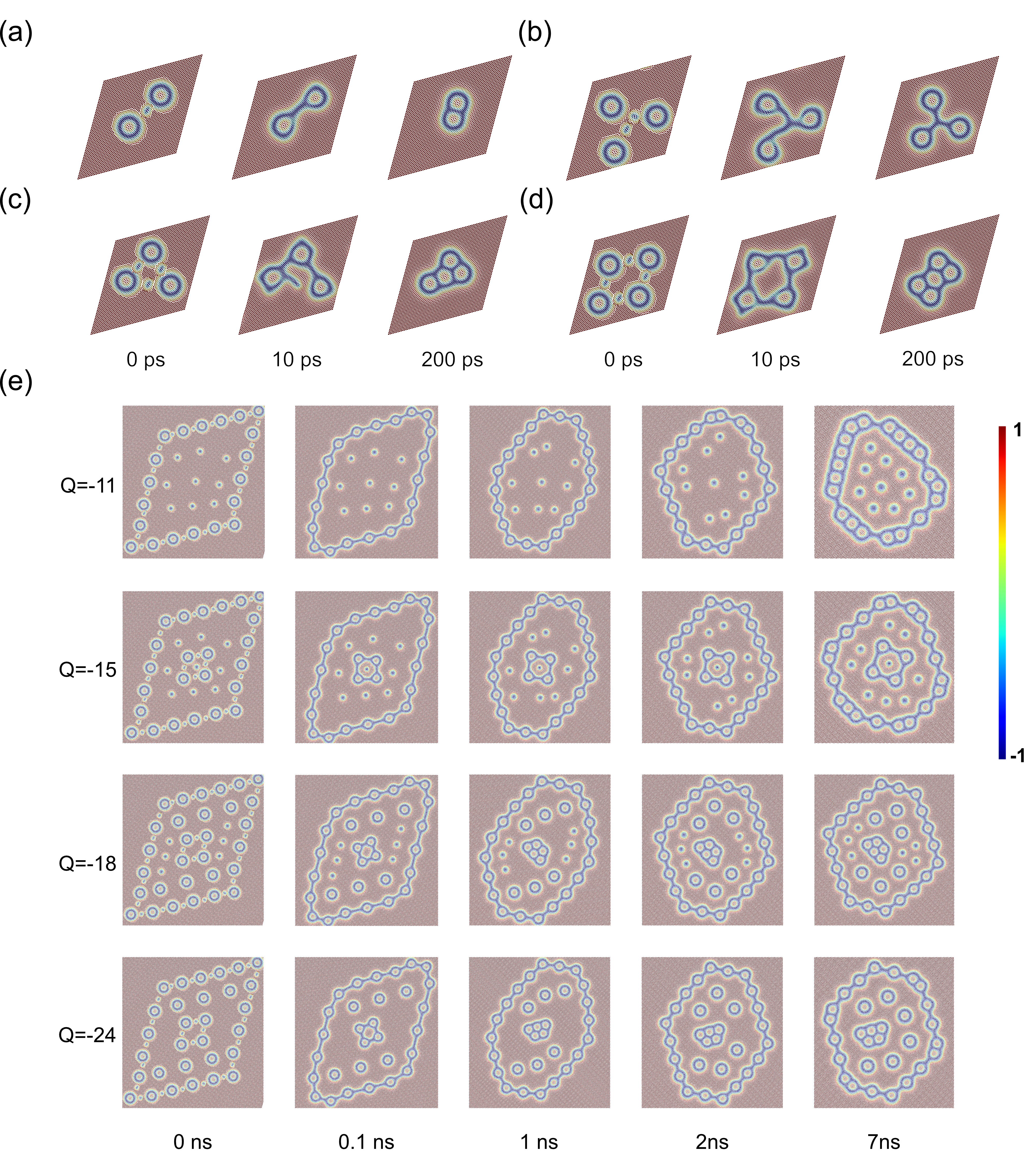}
    \caption{\textbf{Cell-like skyrmionic metamaterials with distinct topological charges.}
    (a-d) skyrmion bags with topological charges ranging from $-1$ to $-4$; 
    (e) skyrmion bags with topological charges ranging from $-11$ to $-24$. Additional details regarding the formation of those textures are available in Supplementary Movie S7.
    The color scheme in the spin configurations is the same as in Fig.\,1.
    }

    \label{fig:fig4}
\end{figure*}

\begin{figure*}
    \centering
    \includegraphics[width=16cm]{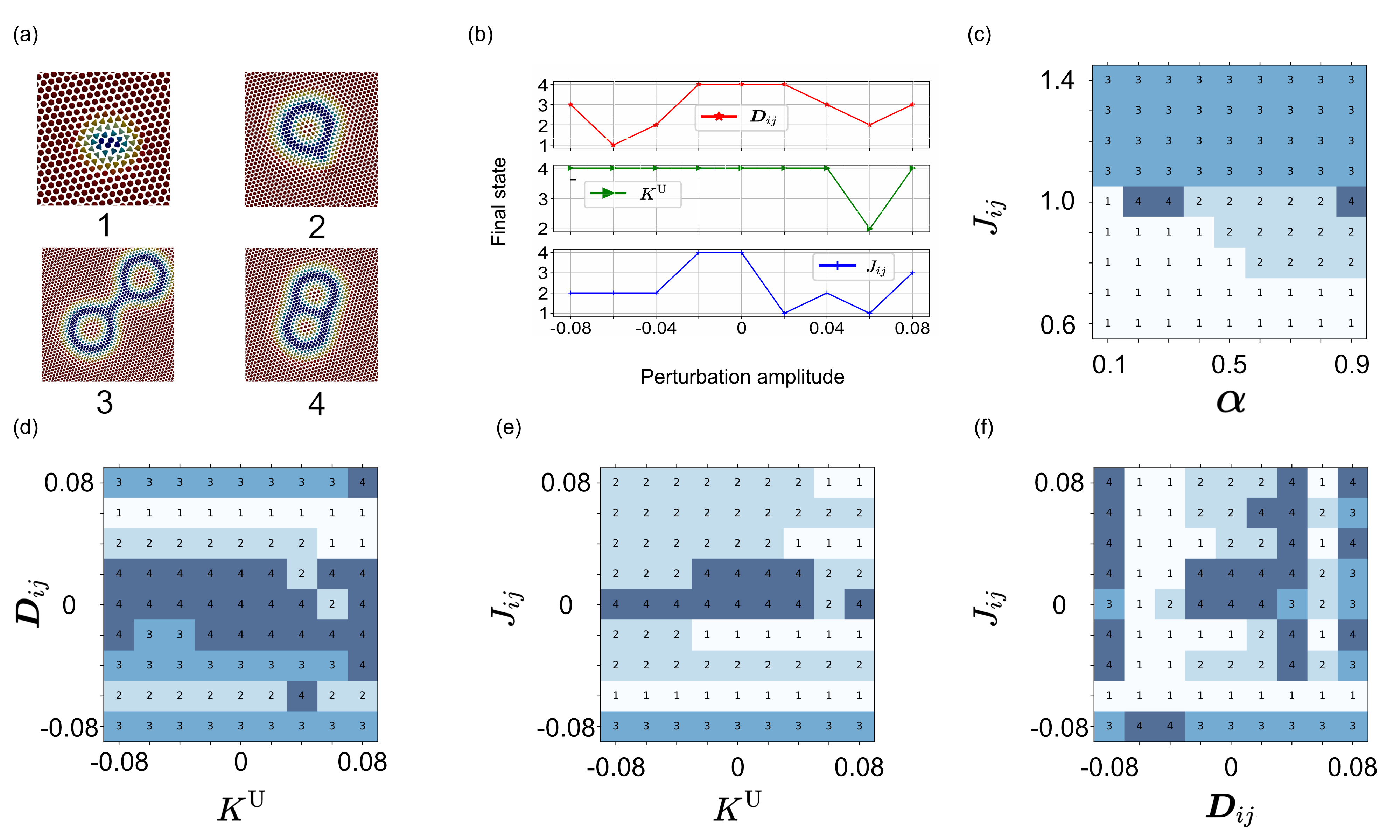}
    \caption{ \textbf{Tests of the robustness and general applicability of the controlled assembly process.} (a) Subplots 1 to 4 display the final states after 200\,ps of an initial state consisting of two skyrmionium and one antiskyrmion (the same for all four subplots). 
     The color scheme is the same as in Fig.\,1.(b) Results under random perturbations of varying maximum amplitudes on each nearest shell
    applied to $\mathbf{D}_{ij}$, $J_{ij}$, and $K^U$.
    (c) Final states after scaling of the exchange interaction parameters $J_{ij}$ and damping $\alpha$ (see Supplementary Note 1), following 200 ps ASD simulations. The color from white to dark blue and the inside number from 1 to 4 indicates the final state displayed in (a).
    (d-f) Final states with random perturbations ($\pm8\%$) applied to combinations of $\mathbf{D}_{ij}$-$K^{\mathrm{U}}$, $J_{ij}$-$K^{\mathrm{U}}$, and $J_{ij}$-$\mathbf{D}_{ij}$, respectively. }
    \label{fig:fig5}
\end{figure*}

\subsection{Flake-like skyrmionic metamaterials}

Together with the aforementioned lattice-like skyrmionic metamaterial, another possible stable type suggested in this work is the \textit{flake}, which is shown in the middle subfigure of Figure \ref{fig:fig3}(c) and (d). To demonstrate the building pathway of a flake, we begin by constructing ring-like prefabricated building blocks ($Q=-6$). As shown on the left of Figure \ref{fig:fig3}(a), according to the site-based topological charge distribution rule, there are two methods for forming this ‘ring’ structure. In the first method (top of Figure \ref{fig:fig3}(a)), six antiskyrmions ($Q=-1$) are positioned close enough, less than the critical distance, to the boundary of a skyrmionium ($Q=0$) to ensure simultaneous attraction. A 300 ps spin dynamics simulation on this initial spin configuration predicts the assembly of the antiskyrmions, forming the target prefabricated building block with a total topological charge of $-6$. The second method (bottom of Figure \ref{fig:fig3}(a)) involves moving an antiskyrmion close to the notch of a crescent-shaped prefabricated building block ($Q=-5$), created from 5 antiskyrmions. In this case, only 100 ps ASD simulations are required to validate the formation of the target high-order antiskyrmion prefabricated building block.

After acquiring this high-order antiskyrmion, we use it as an elementary building block to construct a high-order antiskyrmion lattice, as shown in Figure \ref{fig:fig3}(b). This skyrmionic lattice is predicted to be unstable even at 1 K, where a minor self-rotation can easily move the blocking point, leading to connections between components and unwinding. During this process, we also observe the formation of some flake-like textures.

Expanding on this, we increase the antiskyrmion density based on the previous building block. Ultimately, as shown in Figure \ref{fig:fig3}(c), we isolate a honeycomb flake-like skyrmionic metamaterial ($Q=-22$). In Figure \ref{fig:fig3}(d), by adding one more antiskyrmion, we obtain a honeycomb flake-like skyrmionic metamaterial ($Q=-23$) hosting the well-known 5-7 honeycomb lattice defects\cite{fei2021defective}, where a 5-7 defect in a honeycomb lattice, is a defect in which one ring has 5 sites and an adjacent ring has 7 sites, instead of all rings having 6 sites. Both of those flake-like skyrmionic metamaterials are predicted to be stable even at 10\,K in Pd/Fe/Ir(111). However, the flake with the 5-7 defect is less stable than the one without. To further analyze this, we define the relative defect formation energy as:
\begin{equation}
\begin{aligned}
\label{eq:formation-energy}
E_{rdf} =\frac{1}{n}\left[E_{hf} + n(E_{ASK}- E_{FM}) -E_{df}\right].
\end{aligned}
\end{equation}

Here, $E_{rdf}$, $E_{hf}$, $E_{FM}$, $E_{ASK}$, and $E_{df}$ represent the relative defect formation energy, the energy of the honeycomb flake-like skyrmionic metamaterial under a ferromagnetic background, the energy of the ferromagnetic background, the energy of an isolated antiskyrmion under a ferromagnetic background, and the energy of the honeycomb flake-like skyrmionic metamaterial with a 5-7 defect under a ferromagnetic background, respectively. In Eq. \ref{eq:formation-energy}, $n$ denotes the topological charge difference between the two flakes. Considering the DFT calculated magnetic spin moment of 2.78 $\mu_{B}$/Fe atom, and the applied field $B^{\textnormal{ext}}=3.5$ T, the relative defect formation energy between the two flakes is $E_{rdf}\sim0.15$ mRy/defect. This indicates that introducing defects to the flake-like skyrmionic metamaterial decreases its stability.

\subsection{Stabilized cell-like skyrmionic metamaterial with texture nesting}

To optimize the use of all basic building blocks relevant for this investigation, namely skyrmions, antiskyrmions, and skyrmioniums, as shown in Figure \ref{fig:fig4}(e) in this section we suggest several promising cell-like skyrmionic metamaterials and explore the construction pathway from small skyrmion bags to cell-like skyrmionic metamaterials, inspired by the design of skyrmion bags \cite{PhysRevB.99.064437}. These bags can also be interpreted as magnetic textures stabilized by skyrmionium. 
As demonstrated in Figure \ref{fig:fig4}(a) to (d), we have been able to stabilize several skyrmions with different complexity and topological charges. This involves, e.g., skyrmion bags with topological charge ranging from -1 to -4, by applying a controlled assembly approach to sets of skyrmionium and antiskyrmions. Note that the information in  \ref{fig:fig4} shows the time evolution of the different topological objects. At 0\,ps we show the configuration obtained initially in the controlled assembly protocol, and the data at 200 ps (Fig.\ref{fig:fig4}a-d) shows the final converged configuration.
We observe that in the so conducted, controlled assembly process, the textures inherit an homogeneous, boundary from the skyrmionium, lose the property of attraction but retain their topological charge. Since these skyrmion bags also adhere to the local topological charge distribution, it is expected that more complex skyrmion bags can be constructed through similar operations.

With the introduction of these new kinds of prefabricated building blocks, i.e., skyrmion bags, the repertoire of building blocks with isotropic boundary conditions, or in other words, those that exhibit mutual repulsion, is now expanded beyond the initial three types. Utilizing these building blocks, we introduce a texture nesting operation, which involves using antiskyrmions and skyrmionium to create line-like, closely-packed skyrmion bags as boundaries to nest other textures with isotropic boundary conditions. As shown in Figure \ref{fig:fig4}(e), we conduct a 7 ns ASD simulation in a system of $300\times300$ spins. Due to the large system size and extended simulation time, we set the damping to 0.2 and the time step to 1 fs to expedite convergence. We design a texture, similar to but not identical to one in a previous study\cite{foster2019two}, containing only skyrmions within the skyrmion-bag-ring. This results in a skyrmion bag with a skyrmion lattice inside. Subsequently, we experiment with embedding skyrmionium and small skyrmion bags within the larger skyrmion bag ring, as illustrated in the subsequent rows of Figure \ref{fig:fig4}(e). This controlled assembly mechanism potentially leads to more complex metastable textures on demand. All these dynamic pathways can be summarized as utilizing repulsion from skyrmions and skyrmion bags to inhibit the shrinking process of a newly assembled skyrmion bag.

\subsection{Generalization of results}
\label{sec:Generalization}
Until now, all simulations have been performed using spin Hamiltonian parameters for Pd/Fe/Ir(111) calculated from Density Functional Theory (DFT) \cite{miranda2022band}. These calculations are known to give reliable values of, e.g., the Heisenberg exchange as well as the anisotropic Dzyaloshinskii-Moriya interaction (DMI). However, as with every theory, an uncertainty naturally exists -- in part, due to the approximations involved in all theoretical steps and methods. In order to test the sensitivity of the results predicted here towards such uncertainties, we designed a set of simulations where the magnetic parameters are perturbed by given percentages.  This also extends the controlled assembly protocols to encompass a wider range of materials. To achieve this and address more realistic conditions, we now intend to emulate a possibly common experimental situation: the assembly process occurring among spin textures surrounded by other metastable excitations \cite{Dupe2016,Zhang2017,Hassan2024}. A simplified way to emulate this is to set up a relatively small ($60\times60$) spin lattice with periodic boundaries (PBs). Hence, the set of simulations involves two skyrmioniums and one antiskyrmion in this lattice, and the application of a scalar perturbation to the DFT calculated parameters of Pd/Fe/Ir(111). Four obtained final states (1-4) after a 100 ps ASD simulation at each sampling point are shown in Figure \ref{fig:fig5}(a) subplots 1 to 4. In these simulations, as illustrated in Figure \ref{fig:fig5}(b), we first applied random perturbations ($\pm 8\%$) to the DFT-calculated DMI ($\mathbf{D}_{ij}$), anisotropy ($K^{\mathrm{U}}$), and Heisenberg exchange ($J_{ij}$) interactions. 
 
We found that the final state 4
is not strictly limited to the originally DFT calculated parameters.

Furthermore, to evaluate the general applicability of our findings,
we choose to apply pairwise tuning of the coefficients in the spin Hamiltonian, e.g., the $\mathbf{D}_{ij}$- $J_{ij}$ pair. 
As shown in Figure \ref{fig:fig5}(c), we first scaled the DFT calculated nearest-neighbor $J_{i j}$ from 0.6 to 1.4 and varied the present damping (originally $\alpha=0.1$) from 0.1 to 0.9.  The results indicate that with lower $J_{i j}$ and higher damping, we can successfully form a skyrmion bag. However, increasing $J_{i j}$ leads to more energetic spin wave emissions from the assembly process, which can destabilize structures like skyrmioniums and skyrmion bags compared to antiskyrmions. Interestingly, when $J_{i j}$ is lower, 100 ps may not suffice for the skyrmion bag to transition from a sparse state (as shown in subplot 3 in Figure \ref{fig:fig5}(a)) to a dense state (as shown in subplot 4 in Figure \ref{fig:fig5}(a)). In turn, changes in the $\alpha$ parameter (see Fig. \ref{fig:fig5}(c)) do not directly impact the final states, but instead influence on the lifetimes of spin-wave-like excitations and other transient modes emitted during the assembly process. Thus, for smaller $\alpha$ values these excitations can live longer, facilitating the system to escape shallower metastable states by transiently perturbing them; the probability that the system will stabilize in one of such metastable states decreases, but it is still possible, as shown in Fig. \ref{fig:fig5}(c) for $\alpha=0.2$ or $\alpha=0.3$. This process is corroborated by the finite size of our simulation box with PBs, which eventually causes these excitations to reflect back to the higher-order skyrmionic textures and destabilize them (as in an area surrounded by other topological structures). With this insight, we reverted the Gilbert damping to its original setting ($\alpha = 0.1$) to mitigate the effects of such spin-wave-like excitations. We then analyzed the dynamics of controlled assembly with perturbed interaction parameters ($\mathbf{D}_{i j}$, $J_{i j}$, and $K^{\mathrm{U}}$). 

Similar generalization tests are performed in $\mathbf{D}_{ij}$-$K^{\mathrm{U}}$, $J_{i j}$-$K^{\mathrm{U}}$, and $\mathbf{D}_{ij}$-$J_{i j}$ pairs, as shown in Figures \ref{fig:fig5}(d-f). From these results, it is clear that random perturbations in $K^U$ do not significantly affect the formation of magnetic states. In turn, the behavior under perturbations in $J_{ij}$ and $\mathbf{D}_{ij}$ is more complex: the final state depends, for example, on how the random perturbations affect the frustration in the Fe-Fe exchange interactions (the third-nearest shell in Pd/Fe/Ir(111), see Supplementary Figure 1). Apart from this complex behavior, it is evident that the results are not limited to a fortuitous choice of the Pd/Fe/Ir(111) magnetic interaction parameters, but can survive random perturbations. Skyrmionic metamaterial and controlled assembly are thus anticipated to occur in other systems where long-range frustration and DMI interactions coexist.

\begin{figure*}
    \centering
    \includegraphics[width=17cm]{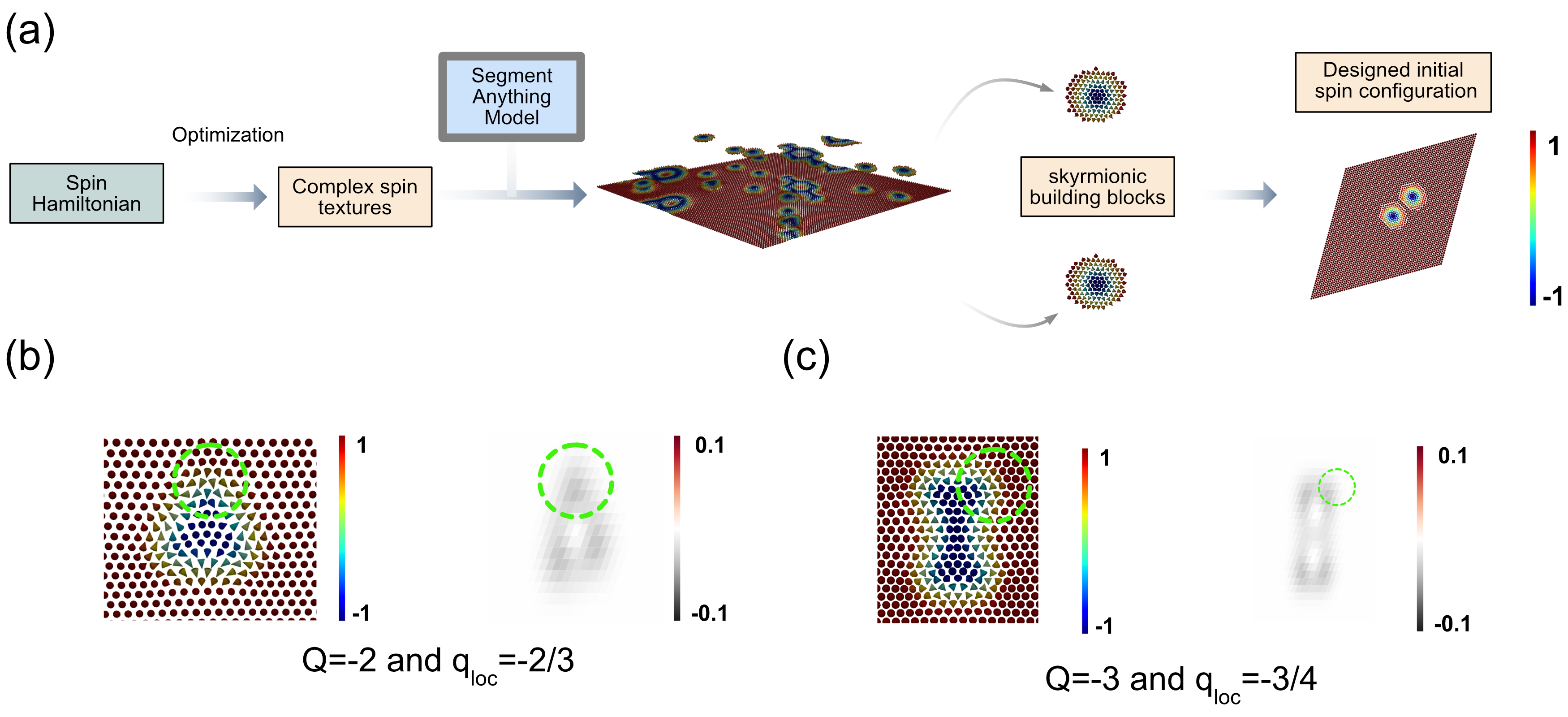}
    \caption{\textbf{Design of initial spin configurations and site-based local topological charge density.}
    (a) Workflow for constructing initial spin configurations for controlled assembly analysis. 
    (b) and (c) Visualization of site-based local topological charge distribution ($q_{\rm loc}$, see Methods). 
    The left panel features real-space visualizations of the magnetic texture, while the right panel depicts the corresponding topological charge density. The color bars ranges are given in terms of the $z$-component for the local normalized spin moment (left panel), and the site-based topological charge density (right panel, unitless). Green dashed circles mark the sites, highlighting the local environment. For the $Q=-2$ structure, the cutoff radius of $r_c\sim0.9$ nm results in $q_{\textnormal{loc}}\sim -\frac{2}{3}$.
    }
    \label{fig:fig6}
\end{figure*}

\section{Discussion}

The ability to use simulation tools to pick up individual atoms and to place them where one wants, so that one can then investigate their possible interaction and stability, has been a fantastic asset in understanding chemical reactions, and therefore deepened the understanding of, e.g., catalysis, crystal growth and surface reconstructions. Several softwares have been designed for this task. The ability to do this with magnetic textures is far less investigated. Here we take steps to accommodate this, and we demonstrate that controlled assembly protocols, that combine artificial intelligence based 
segmentation tools with atomistic spin-dynamics simulations, offer a great opportunity for such investigations.
In this work we began by systematically analyzing the pairwise interactions between skyrmionic elements, pinpointing their critical distances. This understanding facilitates the exploration of single high-order antiskyrmion formation through controlled assembly, guided by the distribution of site-based local topological charges. 
Although manipulating skyrmionic textures experimentally remains challenging, the controlled assembly method presents new potential solutions for constructing complex spin textures. Crucially, we successfully identified a variety of skyrmionic metamaterials. These structures are notable for their stability for a wide range of different Heisenberg and DM interactions, presenting new realms of complexity and stability in magnetic textures. The interactions used for the majority of the investigations presented here, represent a system that has been realized experimentally: the Pd/Fe/Ir(111) system. Hence, the spin Hamiltonian used here is appropriate for a system that has been observed in nature, as opposed to a model Hamiltonian where the interaction parameters may be tuned or fitted, having a specific finding or observation in mind. Judging from the results presented here, the Pd/Fe/Ir(111) system has a fantastic richness in it, in terms of the topological magnetic structures it can host. 

Although the specific capabilities of identified skyrmionic metamaterials, as compared to conventional skyrmionic textures, require further elucidation, the findings of this study significantly contribute to the exploration of the theoretical landscape of skyrmionic textures. Specifically, our research reveals a new set of long-lived textures and their formation pathways, which could potentially act as novel candidate information carriers in the construction of spintronic devices. This work opens promising avenues for the practical realization of long-lived complex magnetic textures, i.e., constructing these textures from fundamental skyrmionic configurations.  We anticipate that our work will catalyze further experimental explorations, thereby advancing the development of future spintronic devices.

\section{Methods}
\label{sec:method}
We describe this system using a classical atomistic spin Hamiltonian, with all relevant interaction parameters derived from Density Functional Theory (DFT) \cite{miranda2022band,xu2023metaheuristic}:

\begin{equation}\label{eq:param-hamiltonian}
\begin{aligned}
\mathscr{H}
&=
-\sum_{i \neq j} J_{i j} \mathbf{S}_{i} \cdot \mathbf{S}_{j}-
\sum_{i \neq j} \mathbf{D}_{i j}\cdot\left(\mathbf{S}_{i} \times \mathbf{S}_{j}\right) \\&- \sum_{i} \mu_{i}\mathbf{B}^{\mathrm{ext}} \cdot \mathbf{S}_{i}  - \sum_{i}K^{\mathrm{U}}_i\left(\mathbf{S}_{i} \cdot \mathbf{e}_{z}\right)^{2},
\end{aligned}
\end{equation}

\noindent where $\mathbf{S}_{i}$ represents the normalized spin moment at site $i$, and $\mu_i$ its magnetic moment length. The parameters $J_{i j}$, $\mathbf{D}_{i j}$, $K^{\mathrm{U}}$, $\mathbf{e}_{z}$, and $\mathbf{B}^{\mathrm{ext}}$ correspond, respectively, to Heisenberg exchange interactions, Dzyaloshinskii–Moriya interactions, uniaxial anisotropy, the easy axis vector, and the applied magnetic field. The DFT calculated parameters are scaled and perturbed in the generalization section using preset factors and restricted Gaussian noise (see Supplementary Note 1). This is done to enhance the generalizability and robustness of the model. The stability test, involving perturbation of input parameters, demonstrates that our findings are not limited to the Pd/Fe/Ir system but also applicable to a broader range of skyrmion-hosting thin-film materials. While dipole-dipole interactions (DDI) are known to stabilize topological skyrmionic structures \cite{Jena2020,Hassan2024,Lobanov2016}, including an explicit DDI term in the spin Hamiltonian (Eq. \ref{eq:param-hamiltonian}) had minimal impact on the results (see discussion in Supplementary Note 3).

Using the spin Hamiltonian of Eq. \ref{eq:param-hamiltonian}, we performed high-throughput ASD simulations in a wide range of initial spin configurations. The ASD simulations were carried out in for a single fcc(111) monolayer (i.e., a hexagonal lattice) representing the Fe layer in the ideal fcc-Pd/Fe/Ir(111) structure, with the Ir experimental lattice parameter (Fe-Fe equilibrium distance of 0.272 nm). The initial spin configurations, where the skyrmionic textures are separated by various distances, can be used to   understand their interactions. To facilitate the construction of systems involving topological structures separated by controlled distances, as shown in Figure \ref{fig:fig6}(a), we employed a state-of-the-art artificial intelligence segmentation tool to extract elemental spin textures from the optimized larger spin system \cite{kirillov2023segment,xu2023metaheuristic}. This tool allows for the systematic design of initial spin textures, such as selecting different element textures, altering the distances between elements, and rotating element textures. We studied the motion of skyrmionic textures at $T = 1$\,mK using atomistic spin dynamics, numerically solving the stochastic Landau–Lifshitz–Gilbert equation as implemented in the Uppsala Atomistic Spin Dynamics (UppASD) package \cite{UppASD_book,Skubic2008}:

\begin{equation}
\begin{aligned}
\label{eq:llg-equation}
\frac{d \mathbf{S}_i}{d t}= -\gamma_{\mathrm{L}} \mathbf{S}_i \times\mathbf{B}_i^{\textnormal{eff}}(t)-\gamma_{\mathrm{L}} \alpha\mathbf{S}_i \times\left(\mathbf{S}_i \times\mathbf{B}_i^{\textnormal{eff}}(t)\right).
\end{aligned}
\end{equation}

In Eq. \ref{eq:llg-equation}, $\alpha$ is the scalar Gilbert damping constant, $\gamma_L=\frac{\gamma}{\left(1+\alpha^2\right)}$ is the renormalized gyromagnetic ratio, and $\mathbf{B}_i^{\textnormal{eff}}(t)$ is the effective magnetic field at the site $i$:

\begin{equation}
\begin{aligned}
\label{eq:effective-mag-field}
\mathbf{B}_i^{\textnormal{eff}}(t)=\mathbf{B}_i+\mathbf{B}_i^{\mathrm{f}}(t)=-\frac{1}{\mu_i}\frac{\partial \mathscr{H}}{\partial \mathbf{S}_i}+\mathbf{B}_i^{\mathrm{f}}(t),
\end{aligned}
\end{equation}

\noindent which here includes a time-dependent stochastic thermal field $\mathbf{B}_i^{\mathrm{f}}(t)$. The random field $\mathbf{B}_i^{\mathrm{f}}$ is assumed to be a Gaussian stochastic process that satisfies the time-averages $\langle B_i^{\mathrm{f},\alpha}(t)B_j^{\mathrm{f},\beta}(t^{\prime})\rangle=2D_i\delta_{ij}\delta_{\alpha\beta}\delta(t-t^{\prime})$ and $\langle B_i^{\mathrm{f},\alpha}\rangle=0$, where $\alpha,\beta$ are Cartesian coordinates, $\delta_{ij}$ is the Kronecker delta, $\delta(t-t^{\prime})$ is the Dirac delta, and $D_i=\alpha k_B T /(\gamma\mu_i)$ is the strength of the thermal fluctuations, determined by the fluctuation-dissipation theorem  \cite{UppASD_book}.

As depicted in Figure \ref{fig:fig6}(b) and (c), we propose a protocol to guide the construction of all skyrmionic metamaterials in 2D films, utilizing a site-based topological charge distribution method. In this context, a 'site' is a qualitative idea that refers to a representative piece of the texture, characterized by a unique local topological charge density distribution. A total of four distinct sites have been defined in this work, with more detailed information available in the Supplementary Note 2.

The topological charge $Q$ of a given continuous two-dimensional magnetization texture with unitary local direction $\vec{n}(x,y)$ is computed using

\begin{equation}
\label{eq:topological-charge-continuum}
Q =\frac{1}{4\pi}\int_A \vec{n}\cdot\bigg(\frac{\partial\vec{n}}{\partial x}\times  \frac{\partial\vec{n}}{\partial y}\bigg)dxdy,
\end{equation}

\noindent where the integral is taken over the area $A$ in the $xy$-plane. Similarly, for a discrete spin system, one can use the Berg and Lüscher invariant \cite{Berg1981}, which reads

\begin{equation}
\label{eq:discrete-topological-charge}
Q = \sum_{x^*}q(x^*),
\end{equation}

\noindent for all unique topological densities $q(x^*)$ attached to vertices $x^*$ in the lattice with spins sitting around, also taking into account the periodic boundary conditions. By construction, both Eqs. \ref{eq:topological-charge-continuum} and \ref{eq:discrete-topological-charge} result in the \textit{total} topological charge existent in the system. The discrete format allows to construct the concept of a \textit{local topological charge}, $q_{\textnormal{loc}}$, which can be defined as the following summation:

\begin{equation}
\label{eq:q-loc}
q_{\textnormal{loc}} = \sum_{x^*\in \{ x \mid r(x) \leq r_c \}}q(x^*),
\end{equation}

\noindent where $r(x)$ denotes the distance of the point $x$ from a reference point in the lattice, and $r_c$ represents a cutoff radius. In this case, obviously $\lim_{r_c\rightarrow\infty}q_{\textnormal{loc}}=Q$.

\section{Data availability}
The interaction parameters used to parameterize the spin Hamiltonian  are made fully available in Zenodo \cite{Xu2025}. Further details and explanations of the data are available from the authors upon reasonable request.

\section{Code availability}
The code used to apply atomistic spin dynamic simulation can be found at https://github.com/UppASD/UppASD, and the code used for 2D spin texture segmentation can be found at: https://github.com/facebookresearch/segment-anything. The code for the visualization can be found at:https://github.com/MXJK851/SpinView/\cite{xu2023spinview}.

\section{acknowledgments}
Fruitful discussions with Filipp N. Rybakov (Uppsala University), Manuel Pereiro (Uppsala University) are acknowledged. We also thank Liuzhen Yang (Stockholm University) and Qinda Guo (Stanford University)
for the peer review teamwork.
Financial support from the Swedish Research Council (grant numbers VR 2016-05980, VR 2019-05304, 2022-04720 and 2023-04239), and the Knut and Alice Wallenberg Foundation (grant numbers 2018.0060, 2021.0246, and 2022.0108) is acknowledged.
A.D. and O.E. acknowledge support from the Wallenberg Initiative Materials Science for Sustainability (WISE), funded by the Knut and Alice Wallenberg Foundation (KAW). 
Q.X. acknowledges financial support from the China Scholarship Council (201906920083). I.P.M.
acknowledges support from the Crafoord Foundation (Grant
No. 20231063). The computations were enabled by resources provided by the National Academic Infrastructure for Supercomputing in Sweden (NAISS) and the Swedish National Infrastructure for Computing (SNIC) at NSC and PDC, partially funded by the Swedish Research Council through grant agreements No. 2022-06725 and no. 2018-05973. GPU resources are provided by KAW (Berzelius-2022-141).

\section{Author Contributions}
Q.X. initiated the idea, devised the project, and designed all simulations. Z.S., A.E., and I.P.M. contributed to all scientific aspects, with significant input from Z.S. in code development, A.E. in providing expertise on physical insights, and I.P.M. in performing all DFT calculations and validating the spin dynamics simulation results. 
O.E. and A.D. secured funding support and managed the project. Q.X, Z.S., A.E., I.P.M., Z.L., A.B, D.T., O.E., W.Y., and A.D. contributed to results discussions, writing, and revision of the manuscript.

\section{Competing Interests}

All authors declare that they have no conflicts of interest.

\bibliography{SA}

\end{document}